\documentclass[12pt,a4paper,tightenlines,nofootinbib,showkeys]{revtex4}

\def\beq{\begin{equation}}
\def\eeq{\end{equation}}
\def\bea{\begin{eqnarray}}
\def\eea{\end{eqnarray}}

\def\FF{F_{a b} F^{a b}}

\def\dd{{\rm d}}
\def\e{{\rm e}}

\def\pt{\partial}
\def\ka{\kappa}

\def\ph{\phi}

\def\LA{\Lambda}

\def\scr{\scriptstyle}
\def\scrscr{\scriptscriptstyle}

\def\W#1{^{\raise2pt\hbox{$\scrscr#1$}}} \def\Y#1{^{\raise2pt\hbox{$\scr#1$}}}
\def\X#1{_{\lower2pt\hbox{$\scrscr#1$}}} \def\Z#1{_{\lower2pt\hbox{$\scr#1$}}}
\def\sqr#1#2#3{{\vbox{\hrule height.#2pt \hbox{\vrule width.#2pt
height#1pt\kern#1pt\vrule width.#2pt}\hrule height.#2pt}\hbox{\hskip.#3em}}}

\begin{document}

\title{Series solutions for a static scalar potential in a Salam-Sezgin Supergravitational hybrid braneworld.}

\author{Benedict M.N. Carter and
Alex B. Nielsen}

\email{bmc55@student.canterbury.ac.nz}\email{ abn16@student.canterbury.ac.nz}

\altaffiliation{Physics and Astronomy Department, University of Canterbury.}

\date{\today}

\begin{abstract}

The static potential for a massless scalar field shares the essential features of the scalar gravitational mode in a tensorial perturbation analysis about the background solution. Using the fluxbrane construction of \cite{CNW1} we calculate the lowest order of the static potential of a massless scalar field on a thin brane using series solutions to the scalar field's Klein Gordon equation and we find that it has the same form as Newton's Law of Gravity. We claim our method will in general provide a quick and useful check that one may use to see if their model will recover Newton's Law to lowest order on the brane.

\end{abstract}


\maketitle

\section{Introduction}

It has long been a dream to derive the properties of our
four-dimensional world from the symmetries of some higher
dimensional spacetime.  Randall-Sundrum models use warped metrics
in five dimensions to obtain a low-energy effective Newtonian
potential for four-dimensional gravity.  Two models have been
proposed: Randall-Sundrum I \cite{RS1} and Randall Sundrum II
\cite{RS2}. Randall-Sundrum I contains two branes in a compact
spacetime and Randall-Sundrum II contains only one brane embedded
in a five-dimensional non-compact Anti de-Sitter spacetime. However, while the
model met with a lot of interest, for example it may provide an
explanation for the large energy difference between the
weak-unification scale and the Planck scale, it has also led to
some problems. It can be shown for example that under general
assumptions the compact Randall-Sundrum I model must contain a
negative tension brane \cite{leblond-myers-winters}. It has been
pointed out that such negative tension branes violate all the
standard energy conditions \cite{Barcelo:2000ta}. Embedding black
holes in Randall-Sundrum II-type models has also led to some
problems, although the tension of the single brane can now be strictly
positive. The Einstein equations evaluated across the brane forbid simple
black holes in our universe (on the brane) and black strings
extending off the brane develop Gregory-Laflamme instabilities at
the AdS horizon and are unstable far from the brane \cite{CHR}.

In six (and higher) dimensions however, it is possible to have
solutions that contain only positive tension branes.  It may even be
possible to have stable black hole solutions in six-dimensions
\cite{louwil}. Here we present an example of a six-dimensional model
based on a supersymmetric fluxbrane solution of  \cite{louwil} and
\cite{GW} and give a general argument to show that its low energy
limit is four-dimensional Newtonian gravity. The model contains a
single brane embedded in a six-dimensional spacetime. Scalar and
electromagnetic fields propagate in the bulk, and we impose regular
closure of the geometry off the brane.

The basic idea is to investigate the behavior of gravity by
approximating the full tensorial analysis with a scalar field
argument. To do this we solve the Klein-Gordon equation on the
background spacetime solution. The background solution must obey
certain boundary conditions, in particular the Einstein equations
across the brane must hold.  Solving this equation allows us to
derive a static potential for the Newtonian limit which should qualitatively
resemble that of the observed universe.

Further details will appear in \cite{CNW1}.

\section{The Model}

We start with the metric (which equates to the bosonic sector of
the model considered in \cite{SS}),
\beq
S= \int_{M}\dd^6 x \sqrt{-g}\left({{\cal R}\over4\ka^2}-\frac{1}{4}
\pt_\mu \ph \pt^\mu \phi  -\frac{1}{4} \e^{- \kappa \phi} \FF
- {\LA\over2\ka^2}\e^{\ka\ph} \right).
\label{action}
\eeq
The metric assumption we use is
\beq ds^2= \Delta(r) d\theta^2 + \frac{r^2}{\Delta(r)}dr^2 + r^2
\overline{g}_{ij}d{x}^{i}d{x}^{j}\,, \eeq
where $\overline{g}_{ij}$ will, for simplicity, be taken to be the
Minkowski metric. We consequently find solutions of the form \cite{Aghababaie:2003ar}
\beq \Delta(r) =
\frac{A}{r^2}-\frac{B^{2}}{r^6}-\frac{1}{8}\Lambda r^2. \eeq
For later use we define $r_{\pm}$ to be the two zeroes of $\Delta$ (there are only two solutions for $0\leq r \, \epsilon \, \mathbf{R}$. %\mathbb not working
To this background fluxbrane geometry we add a single brane and
interpret our four-dimensional universe to be restricted to the
brane in the usual manner.  Note that in general both the scalar
field $\phi$ and the Maxwell field will exist off the brane. Further
details of the construction will appear in \cite{CNW1}.

In order to consistently reproduce our four-dimensional universe on the brane, the gravitational interaction on the brane must reduce, at low energies, to the observed Newtonian potential.  The basic idea here is to use a massless scalar field to model the
behavior of gravity.  Therefore we look for solutions to the
massless Klein-Gordon equation
\beq \nabla^{2}G_{\Phi} =
\frac{1}{\sqrt{-g}}\partial_{\mu}\left(\sqrt{-g}g^{\mu\nu}\partial_{\nu}G_{\Phi}\right)
=\frac{\delta(r-r') \delta(\theta-\theta')
\delta(\mathbf{x}-\mathbf{x}')}{\sqrt{-g}}\,, \label{KleinGordon}
\eeq
with boundary conditions
\bea
G_\Phi|_{r \rightarrow r_-}&<& \infty \,, \label{regular BC} \\
\partial_r G_\Phi |_{r=r_*} &=& 0\,.
\label{neumann BC}
\eea

\section{The Solution}

To simplify the analysis of (\ref{KleinGordon}) we perform a change
of variables to put the differential equation into Sturm-Liouville
form with the variable $\rho$.
\beq \rho=e^{\int (r^{4}\Delta(r))^{-1} dr} \eeq
which becomes
\beq \rho = \frac{r^{4}-r_{-}^{4}}{r_{+}^{4}-r^{4}}\geq0\,, \eeq
where $r_{+}$ and $r_{-}$ are the two roots of $\Delta(r)$. We
also perform the Fourier decomposition of the Green's function
\beq
G_{\Phi}=\int\frac{d^{4}k}{(2\pi)^{5}}e^{ik_{\mu}(x^{\mu}-x'^{\mu})}
\sum^{\infty}_{n=-\infty}e^{in(\phi-\phi')}y_{q,n}(\rho,\rho').
\eeq
This gives a differential equation of the form
\beq\label{diffeqnrho}
\partial_\rho \left( \rho \partial_\rho {y_{q,n}} \right)-\frac{\overline{n}^2}{\rho} \left(\frac{\rho +\frac{r_-^4}{r_+^4}}{\rho+1}\right)^2  {y_{q,n}} +   \frac{\overline{q}^2}{\left(\rho+1\right)^2}  {y_{q,n}} =  \delta(\rho-\rho')\,,
\eeq
where
\bea
q^2&=&-k_\mu k^\mu\,,
\\
\overline{q}^2&=&q^2\frac{\Lambda}{8}(r_+^4-r_-^4)^2\,,
\\
\overline{n}^2&=&n^2r_+^8\,.
\eea
An obvious way to investigate the behavior of solutions to
this differential equation is to use the method of Fr\"{o}benius
and expand the solution out as a power series.  Provided the $r$
coordinate is close to $r_{-}$ the value of $\rho$ will be small.
The two linearly independent solutions are
\beq y_{1}=\sum^{\infty}_{k=0}a_{k}\rho^{k+c_1} \eeq
\beq
y_{2}=\left\{\begin{array}{ll}
\sum^{\infty}_{k=0}b_{k}\rho^{k+c_2} & n\neq0\\
\\
\log(|\rho|)y_1+\sum^{\infty}_{k=0}b_{k}\rho^{k} & n=0 \,.
\end{array}\right.
\eeq
where $c_1$ and $c_2$ are the solutions of the indicial equation for
(\ref{diffeqnrho}), \beq c^2-(n r_-^4)^2=0. \eeq If we impose the
requirement that the solution be regular as $\rho\rightarrow 0$,
then we must set the coefficient of the second independent solution
to $0$ and we are just left with the first solution. The Einstein
equations at the brane also require a Neumann-type boundary
condition for the function $G_{\Phi}$ at the brane.  This, along
with the usual braneworld matching conditions that the metric should
be continuous at the brane and its first derivative should just be a
step function (the stress-energy tensor contains a delta function
source due to the brane,) result in the on-brane solution at
$\rho=\rho_*=r_-/r_+<1$,
\beq y_{q,n}=\frac{y_{1}(\rho_*)}{\rho_* \partial_{\rho}y_{1}(\rho_*)} \label{yqn} \eeq
The $n=0$ mode represents the lowest order of the potential. For
$n=0$ we find \beq y_1(\rho_*)=\sum_{k=0}^\infty  \left((-1)^k
\sum_{j=0}^k e_{j,k} q^{2j}\right) \rho_*^k\,, \eeq where
$e_{0,0}=1$, $e_{0,k}=0 $ for all $ k >0 $ and $ e_{j,k} \geq 0 $
for all $ j>0$ and $k>0$. In the case where this double summation
converges absolutely then all rearrangements converge to the same
limit and we can write \beq y_1(\rho_*)=\sum_{j=0}^\infty
\left(\sum_{k=j}^\infty (-1)^k e_{j,k} \rho_*^k \right) q^{2j}. \eeq
Similarly \beq \rho_* \partial_{\rho}y_{1}(\rho_*)=\sum_{j=0}^\infty
\left(\sum_{k=j}^\infty (-1)^k k e_{j,k} \rho_*^k \right) q^{2j}.
\eeq
We can then write out (\ref{yqn}) as \beq
y_{q,0}(\rho_*)=\sum^{\infty}_{k=-1}Q_{k,0}q^{2k} \eeq and calculate
$Q_{i,0}$ ($Q_{i,n}$ is the general term) order by order in $q^2$
using the relationship \beq y_1(\rho_*)=[y_{q,0}(\rho_*)][ \rho_*
\partial_{\rho}y_{1}(\rho_*)]\,. \eeq
For example,
\beq
Q_{-1,0}=-\frac{1+\rho_*}{\rho_*}\,.
\eeq
Substituting the series back into the expression for the potential we find
\beq
V_{\Phi}=\int\frac{d^{3}\bf{k}}{(2\pi)^{5}}e^{-i\bf{k}.(\bf{x}-\bf{x'})}
\int^{\infty}_{-\infty}\frac{dk^{0}}{i(k^{0}-i\epsilon)}
\left(\frac{Q_{-1,0}}{q^{2}}+\sum^{\infty}_{j=0}Q_{j,0}q^{2j}\right).
\eeq
The term proportional to $q^{-2}$ just gives a contribution of
\beq
V_{\Phi}=\frac{Q_{-1,0}}{2\pi}\int\frac{d^{3}\bf{k}}{(2\pi)^{3}}\frac{e^{-i{\bf
k}.(\bf{x}-\bf{x'})}}{\bf{k}^{2}} \eeq
and thus the potential to lowest order becomes
\beq V_{\Phi}=-\frac{1+\rho_*}{8\pi^{2}\rho_*}\frac{1}{|\bf{x}-\bf{x'}|}\,,
\eeq
which has the desired $1/\bf{x}$ dependence of the familiar Newtonian
gravitational potential.

\section{Conclusion}
The basic model outlined here is capable of reproducing Newton's law of gravity at low energies, and our method demonstrated here provides a method for quickly checking this. Our method does not in general allow one of easily calculate the correction to the lowest order of the static gravitational potential on the brane. Reproduction of Newton's law of gravity to lowest order seems to be a generic
property of (6D) models constructed in this manner.  Several other
models have achieved the same result although often without the
added feature of regularity in the bulk and with added
restrictions on the brane \cite{Aghababaie:2003ar}.  It is the regular closure of the bulk geometry at the totally geodesic submanifolds (bolts) that gives hope to the idea that perturbations of black holes and gravitational waves will not grow without limit and thus further study should be carried out in this direction.
\section{Acknowledgements}
We thank Dr. David Wiltshire for introducing us to this problem and for his supervision.
This research was supported in part by the Marsden Fund administered
by the Royal Society of New Zealand.

\end{document}